\documentclass[JHEP,12pt]{article}
\usepackage{subcaption}
\usepackage{amssymb,amsmath}
\usepackage{xcolor}
\usepackage{braket}
\usepackage{jheppub}
\usepackage{endnotes}
\usepackage{comment} 
\usepackage{relsize}

\usepackage[normalem]{ulem}
\definecolor{block-gray}{gray}{0.85}

\newcommand\cO{\mathcal{O}}

\author[1]{Thomas Hartman,}
\author[2]{Dalimil Maz\'{a}\v{c},}
\author[3]{David Simmons-Duffin,}
\author[\: 4]{Alexander Zhiboedov} 

\affiliation[1]{Department of Physics, Cornell University, Ithaca, NY, USA}
\affiliation[2]{School of Natural Sciences, Institute for Advanced Study, Princeton, NJ 08540, USA}
\affiliation[3]{Walter Burke Institute for Theoretical Physics, Caltech, Pasadena, CA 91125, USA}
\affiliation[4]{CERN, Theoretical Physics Department, CH-1211 Geneva 23, Switzerland}

\emailAdd{hartman@cornell.edu}
\emailAdd{dmazac@ias.edu}
\emailAdd{dsd@caltech.edu}
\emailAdd{alexander.zhiboedov@cern.ch}

\begin{document}

\title{Snowmass White Paper:\\ The Analytic Conformal Bootstrap}

\abstract{
The analytic conformal bootstrap is an array of techniques to characterize, constrain, and solve strongly interacting quantum field theories using symmetries, causality, unitarity,  and other general principles. In the last decade, bolstered by the development of new Lorentzian methods, it has been used to solve conformal field theories at large spin; to place bounds on energy distributions, event shapes, operator product coefficients, and other observables; and to understand aspects of quantum gravity in anti-de Sitter space. We review these advances and highlight several promising areas for future exploration. Targets include developing new methods to close the gap between numerical and analytic bounds, extending the bootstrap beyond conformal fixed points, applications to quantum gravity and cosmology, and building on ties to condensed matter theory and mathematics.
}

\maketitle

\section{Introduction}

Conformal field theories (CFTs) are central to some of the deepest questions in theoretical physics: How can we describe and solve strongly-coupled systems? What are the building blocks of quantum field theories? What is the space of UV-complete theories of quantum gravity? By studying CFTs, we gain a powerful unifying perspective on all of these problems.

At the same time, these deep questions are hard to address with traditional perturbative methods. The {\it conformal bootstrap\/} seeks to constrain and solve CFTs using nonperturbative structures like symmetry, causality, and unitarity. These nonperturbative conditions combine into a surprisingly rigid framework, with wide-sweeping consequences. In the '80s, bootstrap methods were applied by Belavin, Polyakov, and Zamolodchikov to exactly solve an infinite class of 2d CFTs \cite{Belavin:1984vu}. More recently, beginning with the work of Rattazzi, Rychkov, Tonni, and Vichi in 2008 \cite{Rattazzi:2008pe}, numerical investigations of bootstrap conditions have established their power for constraining CFTs in higher dimensions. For example,  bootstrap constraints fix the critical exponents of the 3d Ising model to high precision \cite{El-Showk:2012cjh,El-Showk:2014dwa,Kos:2014bka,Simmons-Duffin:2015qma,Kos:2016ysd}. These promising results have renewed interest in developing a better {\it analytical\/} understanding of bootstrap constraints.

\subsection*{Motivating questions}

CFTs are quantum field theories (QFTs) that are invariant under the conformal group --- a group that includes rotations and translations, and also rescalings of space and time (dilatations). Conformal symmetry emerges at long distances in a tremendous variety of physical systems, from condensed matter physics, through statistical physics, to string theory, and CFTs describe this universal emergent behavior.

The universality of conformal symmetry at long distances gives a framework for organizing QFTs more generally: we can model a QFT as a renormalization group (RG) flow from a (slightly deformed) CFT at short distances to another CFT at long distances. In this sense, CFTs are the building blocks of QFTs. While this framework does not encompass {\it all\/} known QFTs, it has some important virtues. Firstly, it simultaneously describes both weakly-coupled and strongly-coupled systems (even those with no known Lagrangian description). Secondly, CFTs obey rigorous axioms and possess rich mathematical structures that make them easier to understand than general QFTs.

Among these mathematical structures is the operator product expansion (OPE), an algebra underlying all point-like observables in CFTs. Via the OPE, CFTs become the ultimate algebraic objects --- they are controlled by a rigid algebra, but finding the structure constants of that algebra is equivalent to solving the dynamics of a complex, relativistic, quantum many-body system. Associativity of the OPE gives rise to the crossing symmetry equation, expressing the equality of expansions in two different channels:
\begin{equation}
\begin{gathered}
\setlength{\unitlength}{.55in}
\begin{picture}(7.8,2.3)(0,-0.1)
\linethickness{1pt}
\put(1.7,0.4){\line(1,2){0.3}}
\put(1.7,1.6){\line(1,-2){0.3}}
\put(2,1){\line(1,0){0.8}}
\put(2.8,1){\line(1,2){0.3}}
\put(2.8,1){\line(1,-2){0.3}}
\put(1.2,1){\makebox(0,0){$\mathlarger{\sum}_\cO^{\phantom\cO}$}}
\put(4.5,1){\makebox(0,0){$\mathlarger{\sum}_\cO^{\phantom\cO}$}}
\put(3.9,1){\makebox(0,0){$=$}}
\put(1.65,1.8){\makebox(0,0){\small $1$}}
\put(1.65,0.2){\makebox(0,0){\small $2$}}
\put(3.15,1.8){\makebox(0,0){\small $4$}}
\put(3.15,0.2){\makebox(0,0){\small $3$}}
\put(4.88,1.8){\makebox(0,0){\small $1$}}
\put(4.88,0.2){\makebox(0,0){\small $2$}}
\put(6.1,1.8){\makebox(0,0){\small $4$}}
\put(6.1,0.2){\makebox(0,0){\small $3$}}
\put(5.8,1){\makebox(0,0){\small $\cO$}}
\put(2.4,1.2){\makebox(0,0){\small $\cO$}}
\put(5.5,0.6){\line(0,1){0.8}}
\put(5,0.35){\line(2,1){0.5}}
\put(5.5,0.6){\line(2,-1){0.5}}
\put(5.5,1.38){\line(2,1){0.5}}
\put(5,1.65){\line(2,-1){0.5}}
\end{picture}
\end{gathered}
\label{eq:crossingequation}
\end{equation}
Using associativity and unitarity, analytic bootstrap methods have uncovered important features of the OPE in simplifying limits, such as the large spin limit. The myriad connections between CFTs and physical phenomena mean that these results have wide applications. For example, analytic large-spin calculations can be used to simultaneously understand gravitational bound states in AdS, energy distributions in collider experiments, and cubic symmetry breaking in critical magnets.  Analytic bootstrap methods have also been used to show that certain special OPE algebras saturate universal bounds on scaling dimensions and other observables. An important goal is to further close the gap between numerical bootstrap bounds and analytic bounds.

Via the AdS/CFT correspondence, CFTs are equivalent to UV-complete theories of quantum gravity in Anti de Sitter (AdS) space. Outside of AdS, the question of which UV completions exist is an elusive one. One reason is that we do not know the axioms of quantum gravity in general spacetimes. (Some features of effective field theory should break down --- but which ones?) By contrast, the axioms of CFT are clear and relatively well understood. Remarkably, this makes it possible to reason with confidence about quantum gravity in AdS by studying CFTs. For example, analytic bootstrap methods make it possible to rigorously constrain and study gravitational $S$-matrices in AdS, despite the absence of well-established axioms for gravitational S-matrices in flat space. Analytic bootstrap methods have highlighted the central role of causality and unitarity in constraining UV-completions, and these lessons may have important implications as we seek to understand UV-completions in other contexts, including our own universe.

\section{Recent progress}

\subsection*{Lightcone bootstrap and the large-spin expansion}

There are no known solvable, interacting CFTs in $d>2$ and most of our knowledge about their properties comes from simplifying limits, such as the $\epsilon$-expansion, weak coupling, or large $N$. The lightcone bootstrap reveals simplicity and universality of the CFT data at large spin by analyzing the crossing equation in the lightcone limit.

As a pair of operators becomes light-like separated, the corresponding OPE channel is dominated by the unit operator, with small corrections coming from low twist operators. Crossing symmetry~\eqref{eq:crossingequation} relates proximity to the light-cone in one channel to the inverse spin of operators in the other channel. In this way, the lightcone bootstrap leads to the conclusion that every CFT admits a large-spin expansion, and that CFT operators organize themselves into additive multi-twist families \cite{Alday:2007mf,Fitzpatrick:2012yx,Komargodski:2012ek}.
It allows one to perform systematic analytic computations \cite{Fitzpatrick:2014vua,Alday:2015eya,Alday:2015ewa,Alday:2016njk,Simmons-Duffin:2016wlq}, and estimate properties of double-twist operators even in strongly coupled CFTs, such as the 3d Ising model, when no small parameter is available \cite{Alday:2015ota,Simmons-Duffin:2016wlq}. The same idea can be generalized to many other situations \cite{Bissi:2022mrs}, such as defect CFTs \cite{Lemos:2017vnx}, torus partition functions \cite{Collier:2016cls,Benjamin:2019stq}, thermal correlators \cite{Iliesiu:2018fao,Iliesiu:2018zlz,Alday:2020eua}, event shapes \cite{Chen:2022jhb,Chang:2022ryc}, and higher-point correlation functions \cite{Antunes:2021kmm}.

Ultimately, the lightcone bootstrap and its cousins restate the basic structure of correlation functions dictated by locality and causality in terms of the CFT spectral data. In the simplest case of the four-point function, recent developments put  predictions of the lightcone bootstrap on a firmer theoretical basis by supplementing its predictions with rigorous error bars \cite{Qiao:2017xif,Caron-Huot:2020adz}. For multi-twist operators, this is still an open problem. In any future approach to a nonperturbative solution of CFTs, it is desirable to make the universal large-spin structure as manifest as possible and focus computational efforts on non-universal aspects of the theory.

\subsection*{Inversion formula and the double commutator}

The OPE expresses a four-point correlation function as a discrete sum of conformal blocks, corresponding to exchanges of the physical operators in the theory. There is a more primitive expansion of the four-point function (sometimes called the partial wave decomposition) in terms of a complete basis of orthonormal single-valued functions labeled by integer spin $J \in \mathbb{Z}_{\geq 0}$ and complex dimensions $\Delta = \frac{d}{2} + i \nu$. These correspond to the principal series representations of the Euclidean conformal group \cite{Dobrev:1977qv}.
The usual OPE is recovered from the partial wave decomposition by deforming the $\nu$-contour and picking up poles at the position of the physical operators \cite{Dobrev:1975ru}. It turns out that the coefficients $c(\Delta, J)$ in this decomposition  can be expressed (for $J>1$) in terms of a double commutator integrated over a Lorentzian region \cite{Caron-Huot:2017vep,Simmons-Duffin:2017nub}. This result is known as the Lorentzian inversion formula.

The Lorentzian inversion formula makes manifest many interesting properties of CFT spectra. It implies that local operators and OPE coefficients are organized into Regge trajectories \cite{Costa:2012cb} which are analytic functions of spin that make sense even away from integer values. It is in this precise sense that we can think of spin as an expansion parameter in the light-cone bootstrap. Analyticity in spin ultimately stems from boundedness of correlators in the Regge limit \cite{Maldacena:2015waa,Caron-Huot:2017vep}, a simple but powerful result underlying the dispersive functionals discussed below, and with other important applications, see e.g.\ \cite{Chowdhury:2019kaq,Chandorkar:2021viw}.

 The double commutator that enters the Lorentzian inversion formula is a much simpler object than the full correlator, e.g.\ it suppresses the contribution from double-twist operators, 
yet it captures most of the dynamical information contained in the correlation function. It is also non-negative and bounded from above, which makes it suitable for many bootstrap applications. In the planar limit, the double commutator receives contributions from single trace operators only and thus directly probes the spectrum of one-particle states in the dual AdS theory. This has proven instrumental for loop computations in AdS \cite{Aharony:2016dwx,Alday:2017gde,Aprile:2017bgs,Aprile:2017xsp,Alday:2017vkk,Alday:2017xua}, as well as in perturbative computations \cite{Henriksson:2017eej,Alday:2017zzv,Henriksson:2018myn,Alday:2019clp,Henriksson:2020fqi}.

\subsection*{Energy conditions, light-ray operators, and event shapes}

Local energy is not bounded from below in unitary relativistic QFTs. The same is true for any expectation value of local operators in a bounded region. The situation is different for non-local operators, of which light-ray operators are simple examples. Perhaps the most famous example of a light-ray operator is the integral of the stress-energy tensor over a null geodesic, $\int_{-\infty}^\infty d \lambda T_{\alpha \beta} u^{\alpha} u^{\beta}$. The averaged null energy condition (ANEC) states that this operator is nonnegative. This condition is an input to many fundamental results in general relativity \cite{Borde:1987qr} and CFT \cite{Hofman:2008ar}, but in interacting field theories it was derived only recently. Intriguingly, it was derived by two different methods, one based on the analytic bootstrap \cite{Hartman:2015lfa,Hartman:2016lgu} and the other based on quantum information theory \cite{Faulkner:2016mzt}; both methods have close ties to the emergence of causality in the AdS/CFT correspondence \cite{Kelly:2014mra}. The bootstrap method can also be viewed as a highly boosted limit of the Lorentzian inversion formula.

In the context of CFTs, light-ray operators make a natural appearance when thinking about a conformal collider experiment \cite{Hofman:2008ar}, where they play the role of energy and charge calorimeters placed at null infinity.  More recently, it was realized that generic points on Regge trajectories away from integer spins have a natural interpetation in terms of light-ray operators \cite{Kravchuk:2018htv}. While being non-local, light-ray operators retain commutativity properties of local operators in a nontrivial way \cite{Kologlu:2019bco,Belin:2019mnx}, which leads to interesting sum rules on the CFT data. Lightray operators built out of the stress-energy tensor form the BMS algebra \cite{Cordova:2018ygx}, as well as more general algebraic structures \cite{Huang:2019fog,Belin:2020lsr,Besken:2020snx,Korchemsky:2021htm,Huang:2022vcs}.

Event shapes are observables used to describe hadronic events at colliders. Traditionally, they are computed using scattering amplitudes which requires delicate cancellations of infrared divergencies. In the context of CFTs, event shapes become matrix elements of light-ray operators \cite{Sveshnikov:1995vi,Korchemsky:1997sy,Hofman:2008ar,Kologlu:2019mfz,Korchemsky:2021okt}, and therefore can be efficiently studied using bootstrap techniques in both perturbative \cite{Belitsky:2013xxa,Belitsky:2013bja,Belitsky:2013ofa,Henn:2019gkr} and nonperturbative settings. Remarkably, the machinery of the OPE \cite{Hofman:2008ar,Korchemsky:2019nzm,Kologlu:2019mfz,Chang:2020qpj} and crossing equations  \cite{Chen:2022jhb,Chang:2022ryc} can be generalized to light-ray operators in a nontrivial way. The light-ray OPE has interesting applications in the study of jet substructure in QCD \cite{Dixon:2019uzg,Chen:2020vvp}. Developing a better understanding of the space of light-ray operators and associativity of the light-ray OPE is an important open problem in our quest for understanding nonperturbative Lorentzian dynamics of CFTs.

\subsection*{Dispersion relations and dispersive sum rules}

Conformal dispersion relations express a four-point function in terms of its double commutator \cite{Carmi:2019cub}. They can be derived, for example, by plugging the Lorentzian inversion formula into the conformal partial wave expansion and performing the sum and integral over principal series representations. Alternatively, conformal dispersion relations admit a particularly simple derivation in Mellin space, where they follow from applying Cauchy's theorem to the nonperturbative Mellin amplitude \cite{Penedones:2019tng}. Finally, conformal dispersion relations can be thought of in terms of an expansion in a complete set of functionals with double zeros at double-twist locations \cite{Mazac:2019shk}. 

All of these closely-related constructions lead to so-called ``dispersive sum rules"  \cite{Caron-Huot:2020adz}, which are crossing symmetry constraints expressed in terms of the double commutator. Perhaps the simplest example of a dispersive sum rule is the statement that ANEC operators commute \cite{Kologlu:2019bco}.  In practice, dispersive sum rules possess double zeros at double-twist locations, except for a finite number of Regge trajectories.  In the context of holographic CFTs, double-twist operators are two-particle states in AdS. Dispersive sum rules effectively throw out this redundant two-particle information, focusing on the physics of light particles and the heavy states that provide their UV completion. In the AdS bulk, they connect known IR physics to unknown (but positive) UV data. In \cite{Caron-Huot:2021enk}, dispersive sum rules were constructed that become flat-space dispersion relations in an appropriate limit, allowing one to apply technology from the $S$-matrix bootstrap to constrain holographic CFTs. This leads to detailed bounds on the Wilson coefficients of AdS effective field theories that can be UV-completed, discussed in more detail below. A systematic understanding of the space of dispersive sum rules and lessons that can be learned from them is an important open problem. 

Finally, conformal dispersion relations naturally lead to an expansion of the correlator in so-called Polyakov or Polyakov-Regge blocks. These blocks are closely related to exchange Witten diagrams and arise by applying the dispersion relation to a single conformal block. A similar expansion had been used by Polyakov \cite{Polyakov:1974gs} and more recently in \cite{Sen:2015doa,Gopakumar:2016wkt,Gopakumar:2016cpb,Dey:2016mcs,Dey:2017fab,Gopakumar:2018xqi} as an alternative starting point for the analytic conformal bootstrap. Conformal dispersion relations provide a rigorous underpinning for this idea \cite{Caron-Huot:2020adz,Sinha:2020win,Gopakumar:2021dvg}.

 \subsection*{Analytic functionals and sphere packing}
One of the early successes of the conformal bootstrap was the observation that bounds on the CFT data are often close to being saturated by interesting CFTs, such as the critical 3D Ising model \cite{El-Showk:2012cjh}. The bounds are obtained by numerically solving an optimization problem in a space of linear functionals acting on the crossing equation~\eqref{eq:crossingequation}. This leads to several interesting questions: Do physical CFTs saturate the bounds exactly? If so, is there an analytic proof of the exact saturation? It turns out that at least in some cases, the answer to both questions is yes \cite{Mazac:2016qev}. In these cases, it is possible to construct an analytic functional which certifies the exact saturation \cite{Mazac:2016qev,Rychkov:2017tpc,Mazac:2018mdx,Mazac:2018ycv,Mazac:2018qmi,Hartman:2019pcd,Paulos:2019gtx,Afkhami-Jeddi:2020ezh,Caron-Huot:2020adz,Paulos:2020zxx,Ghosh:2021ruh,Paulos:2021jxx}. In essentially all the examples where an analytic functional has been found, the corresponding solution to the bootstrap is a free theory. Analytic functionals for free theory are examples of dispersive sum rules, discussed above. It is an important open problem to understand whether interacting CFTs exactly saturate bootstrap bounds coming from a small set of correlators, and to construct the requisite analytic functionals.

There is a remarkably close correspondence between the analytic functionals and the recent solution of the sphere packing problem in 8 and 24 dimensions \cite{Hartman:2019pcd}. The key observation is that a method for proving upper bounds on the sphere packing density due to Cohn and Elkies \cite{CE} can be formulated as a bootstrap problem. When the analytic functionals of \cite{Mazac:2016qev} are applied to this problem, they lead to directly to Viazovska's solution \cite{SpherePacking8,SpherePacking24} of sphere packing in 8 and 24 dimension. This establishes a precise connection between the conformal bootstrap and pure mathematics. Other examples of such connections are the recent application of bootstrap techniques to study Laplacian spectra on manifolds \cite{Bonifacio:2020xoc,Bonifacio:2021msa,Kravchuk:2021akc,Bonifacio:2021aqf}, and explorations of 2D CFTs through the lens of quantum codes \cite{Dymarsky:2020bps,Dymarsky:2020qom,Dymarsky:2020pzc,Dymarsky:2021xfc,Buican:2021uyp}. It is likely that many more examples will be found in the future. In this context, a particularly important task is to look for mathematical structures which automatically lead to solutions of the conformal bootstrap equations \cite{Gadde:2017sjg,Kravchuk:2021akc}.

 \subsection*{Constraints on gravitational theories in AdS}
 As explained in the introduction, we can use the conformal bootstrap to study gravitational theories in AdS space. To do so, we first need to provide a general characterization of CFTs dual to weakly coupled gravity in AdS \cite{Heemskerk:2009pn}. The requirement of weak gravitational coupling translates to the CFT having many degrees of freedom per site, i.e.\ large $N$. Furthermore, we are typically interested in gravity theories with no light particles of spin greater than two. In CFT language, this maps to a large lower bound $\Delta_{\mathrm{gap}}$ on the dimension of the lightest higher-spin single-trace operator. This can only be achieved if the CFT is strongly coupled.
 
Equipped with this characterization, we can start exploring the space of CFTs with large $N$ and large $\Delta_{\mathrm{gap}}$ using the conformal bootstrap. Conveniently, in the $N\rightarrow \infty$ limit, any such CFT becomes a mean field theory, and thus can be solved explicitly. One can then expand the CFT data perturbatively in $1/N$ and $1/\Delta_{\mathrm{gap}}$. The perturbative corrections are constrained by the CFT crossing equations. There is a significant amount of evidence that the general solution of this perturbative bootstrap exactly agrees with the predictions of a general effective field theory in AdS \cite{Heemskerk:2009pn,Penedones:2010ue,Fitzpatrick:2010zm,Fitzpatrick:2011ia,Paulos:2011ie,Fitzpatrick:2011dm,Meltzer:2019nbs,Meltzer:2020qbr}. This is analogous to the statement that the low-energy expansion of a general consistent $S$-matrix can be generated using effective field theory in flat space \cite{Weinberg:1978kz}. However, in AdS, it is often much easier to analytically solve the bootstrap equations than it is to evaluate Witten diagrams. This idea has led to the evaluation of a large amount of CFT data in theories with weakly coupled bulk duals \cite{Aharony:2016dwx,Rastelli:2016nze,Rastelli:2017udc,Alday:2017vkk,Aprile:2017bgs,Alday:2017xua,Aprile:2017xsp,Caron-Huot:2018kta,Sleight:2018epi,Sleight:2018ryu,Carmi:2019ocp,Carmi:2020ekr,Carmi:2021dsn}.

Let us reiterate that the conformal bootstrap provides a precise nonperturbative definition of what it means to have a UV complete theory of gravity in AdS. However, by expanding the crossing equations perturbatively in $1/N$ and $1/\Delta_{\mathrm{gap}}$, we are throwing away valuable bootstrap constraints relating the UV and the IR. In other words, the set of fully consistent gravitational theories in AdS could be much smaller than the set of effective field theories one can write down! Thus the conformal bootstrap can provide a rigorous underpinning for the swampland program \cite{Vafa:2005ui,Arkani-Hamed:2006emk}, which aims to identify which effective field theories admit a UV completion; see \cite{Palti:2019pca} for a review.

To describe some recent progress on this front, consider the space of gravitational effective field theories. It can be parametrized by a set of light particles including the graviton and the infinite set of their local interaction vertices. 
It is well known that causality leads to sign constraints on some higher-derivative couplings in quantum gravity \cite{Brigante:2007nu, Hofman:2008ar,Hofman:2009ug,Camanho:2014apa}. This has been derived from the analytic bootstrap \cite{Afkhami-Jeddi:2016ntf,Kulaxizi:2017ixa,Costa:2017twz,Meltzer:2017rtf}, and recently, it has been generalized to prove new strict bounds on higher-derivative couplings in gravitational theories in AdS \cite{Caron-Huot:2021enk}. The idea relies on the  dispersive CFT sum rules reviewed above. These sum rules equate EFT couplings with sums over the ultraviolet states. Unitarity of the UV contributions then leads to bounds on the IR couplings. As a result, the analytic bootstrap sheds light on the emergence of causality in quantum gravity, and helps to constrain the landscape of consistent theories.
 
\subsection*{More general observables}
We have focused so far on vacuum four-point functions of light operators. There are many other observables one can define in a CFT, which lead to additional bootstrap constraints. A simple example is vacuum correlators of $n>4$ local operators. At first sight, these may seem superfluous since crossing symmetry of all four-point functions in principle implies consistency of arbitrary $n$-point functions. However, in practice, higher-point functions may package some CFT data together in a more useful way. For example, they give us access to structure constants of multiple operators with large spin, and multi-trace operators in large-$N$ theories \cite{Rosenhaus:2018zqn,Goncalves:2019znr,Bercini:2020msp,Antunes:2021kmm,Barrat:2021tpn}.

One can also consider correlation functions in the presence of extended objects, such as boundaries, lines, or more general defects. It is natural to demand that the defects preserve the maximal subgroup of the conformal group. In that case, the set of conformal bootstrap constraints can be enlarged to the full set of correlation functions involving local operators distributed in spacetime, as well as on various conformal defects \cite{Cardy:1984bb,Cardy:1989ir,Cardy:1991tv,Liendo:2012hy,Billo:2016cpy}. Various analytic bootstrap methods, such as the large-spin expansion, inversion formulas and analytic functionals can be extended to these situations \cite{Gliozzi:2015qsa,Rastelli:2017ecj,Lemos:2017vnx,Lauria:2018klo,Bissi:2018mcq,Kaviraj:2018tfd,Mazac:2018biw,Liendo:2018ukf,Liendo:2019jpu,Gimenez-Grau:2019hez,Dey:2020jlc,Gimenez-Grau:2021wiv,Barrat:2021yvp}.

Furthermore, correlation functions of a \textit{local} CFT should be consistent not only in flat space, but on an arbitrary manifold. A well known consequence is modular invariance of the partition function of a 2D CFT on $S^1\times S^1$. Famously, it leads to a universal upper bound on the dimension of the first nontrivial Virasoro primary \cite{Hellerman:2009bu}. For recent developments, see \cite{Keller:2012mr,Friedan:2013cba,Collier:2016cls,Anous:2018hjh,Afkhami-Jeddi:2019zci,Hartman:2019pcd,Benjamin:2019stq,Alday:2019vdr,Benjamin:2020mfz}. 

In $d>2$, perhaps the simplest manifold not conformally equivalent to $\mathbb{R}^d$ is $S^1\times \mathbb{R}^{d-1}$. Placing a CFT on this geometry is equivalent to studying its physics at nonzero temperature, or equivalently at fixed energy density. This is also very interesting from the point of view of holography since finite temperature states are dual to black holes. Demanding the consistency of the OPE with periodicity around the thermal circle leads to constraints on finite temperature observables \cite{El-Showk:2011yvt,Iliesiu:2018fao,Iliesiu:2018zlz,Alday:2020eua}. More elaborate geometries also lead to interesting constraints and further connections to black hole thermodynamics \cite{Shaghoulian:2015dwa,Shaghoulian:2015kta,Nakayama:2016cim,Hasegawa:2016piv,Belin:2016yll,Shaghoulian:2016gol,Horowitz:2017ifu,Gadde:2020bov,Giombi:2020xah}. 

Thermal properties of CFTs emerge from average properties of heavy operators. To see this, note that $S^1\times \mathbb{R}^{d-1}$ is a limit of $S^1\times S^{d-1}$ as the radius of the second factor goes to infinity. In this limit, we are sensitive to high-energy states on $S^{d-1}$, or in other words, to local operators with large scaling dimension. In holography, individual heavy operators are dual to black hole microstates. It is a fascinating question to understand the emergence of macroscopic physics from the statistical properties of heavy operators. In the context of the bootstrap, average properties of heavy operators can be probed rigorously using Tauberian theorems \cite{Pappadopulo:2012jk,Mukhametzhanov:2018zja,Mukhametzhanov:2019pzy,Ganguly:2019ksp,Pal:2019yhz,Pal:2019zzr,Mukhametzhanov:2020swe}. Interesting probes of heavy operators also emerge from studying the Regge limit \cite{Caron-Huot:2020ouj}, and higher-point correlation functions \cite{Anous:2021caj}. In CFTs with global symmetries, there exists a similar macroscopic limit, where we take large charge at fixed charge density. Recently, it was realized that this limit admits a description in terms of an effective field theory \cite{Hellerman:2015nra}. It turns out that one can formulate the conformal bootstrap directly in this limit, which reproduces the EFT description, while also allowing for more exotic possibilities \cite{Jafferis:2017zna}. This is yet another instance of the general principle that effective field theory can be thought of as a device for systematically solving the bootstrap equations in a perturbative limit.

\section{Future directions}

\subsection*{Bootstrapping quantum gravity in AdS}

As we reviewed, the study of consistent gravitational theories in AdS is naturally included as a corner of the conformal bootstrap. One of the most important directions for future research is to understand the full set of constraints that the bootstrap imposes on this class of theories. In particular, what are the bounds on the spectra of light particles and their low-energy couplings in the presence of gravity?

An important goal is to test the hypothesis of string universality: Do all weakly-coupled, consistent theories of gravity come from string theory, or are there counterexamples? The analytic conformal bootstrap toolkit is mature enough to begin addressing this question.

One promising approach is to apply dispersive sum rules \cite{Penedones:2019tng,Caron-Huot:2020adz,Gopakumar:2021dvg}. Even for a single four-point function, their consequences are far from fully understood. To make further progress, it will also be essential to generalize dispersive techniques to systems of correlators and higher-point functions.

It is important to stress that consistent gravitational theories satisfy a much larger set of constraints than ordinary quantum field theories. Indeed, the presence of dynamical gravity in AdS implies locality of the dual CFT. Not only do such CFTs contain a stress tensor, but their correlation functions must be consistent on general manifolds, in contrast with theories dual to non-gravitational QFTs in AdS.

The additional constraining power of more general geometries is apparent already in AdS$_{3}$/CFT$_{2}$, where  modular invariance, i.e.\ consistency on the torus, implies a universal upper bound on $\Delta^{\text{2d}}_{\text{gap}}$, the dimension of the lightest nontrivial primary \cite{Hellerman:2009bu}. Conceivably, studying local higher-d CFTs on various manifolds will lead to similarly far-reaching consequences. Staying in the realm of 2d CFTs, what is the optimal upper bound on $\Delta^{\text{2d}}_{\text{gap}}$ at large central charge? In particular, how can we go beyond the current analytical record $\Delta^{\text{2d}}_{\text{gap}} \leq c/8.503$ \cite{Hartman:2019pcd} and even beyond the numerical estimate $\Delta^{\text{2d}}_{\text{gap}} \lesssim c/9.1$  \cite{Afkhami-Jeddi:2019zci} coming from the spinless modular bootstrap?

The task of charting the space of UV-completable gravitational effective field theories is also the goal of the swampland program \cite{Vafa:2005ui,Palti:2019pca}. This program has generated a large set of far-reaching conjectures concerning properties of quantum gravity. These conjectures are supported, to a varying degree of rigor, by physical arguments as well as exhaustive searches of examples in string theory. In contrast, bounds coming from the bootstrap have typically been less dramatic, but benefit from a high degree of rigor since they only rely on unitarity, causality and symmetry. Through a combination of ideas from the two sides, the gap between bootstrap bounds and swampland conjectures will continue to narrow, and should ultimately disappear.

\subsection*{Bootstrapping holographic duality beyond AdS/CFT}
The prototypical examples of holographic duality involve a $d$-dimensional CFT on the conformal boundary of anti-de Sitter spacetime, and in these cases there is a direct map from the analytic bootstrap to properties of quantum gravity. However, holographic duality is believed to extend beyond AdS, and there are many potential insights to be gained by applying bootstrap methods to quantum gravity in other settings. 

One route is to take the flat-space limit of AdS/CFT. This has already been used to develop $S$-matrix bootstrap methods and to derive bounds on gravitational scattering \cite{Caron-Huot:2021rmr,Caron-Huot:2021enk}, but the bounds are not sharp in $D=4$ due to infrared divergences that appear when the AdS size becomes infinite. Can CFT methods be used to construct good IR-safe boundary observables for four-dimensional gravity in asymptotically flat space?

Another perspective on holography in flat spacetime starts from the observation that flat-space scattering amplitudes have CFT-like kinematics on the celestial sphere. This has led to the notion of celestial  holography \cite{Pasterski:2016qvg}. Can the methods of the analytic bootstrap such as functionals, dispersion relations and OPE inversion be applied to celestial CFTs? Additionally, how are the light-ray operators of the analytic bootstrap related to the conservation laws, soft theorems, and memory effects of celestial CFT?

The bootstrap can be applied to observables in de Sitter spacetime, which is relevant to our universe during the early stages of expansion as well as the present-day era of dark energy. There are various proposals for holographic descriptions of de Sitter in which the CFT lives, for example, on the spacelike future boundary, at the edge of a flat-space bubble, or on the de Sitter horizon. How are unitarity and causality manifested in the CFT description? There has been recent progress on this question from various points of view, involving both CFT methods and scattering amplitudes. The most interesting question is perhaps how best to extract constraints on inflationary observables; see e.g.~\cite{Baumann:2015nta,Cordova:2017zej,Melville:2019wyy,Baumann:2019ghk,Ye:2019oxx,Pajer:2020wnj,deRham:2020zyh,Pajer:2020wxk,Grall:2021xxm,Goodhew:2021oqg,Meltzer:2021zin,Hogervorst:2021uvp,DiPietro:2021sjt,Melville:2022ykg} for recent steps in this direction.

\subsection*{Conformal manifolds and deformations}
The QFT landscape has isolated points with conformal symmetry; plateaus parameterized by exactly marginal couplings; and renormalization group flows activated by relevant or irrelevant couplings, or by moving into the moduli space of vacua. On the plateaus --- including the ${\cal N}=4$ SCFT in four dimensions which features in the AdS/CFT correspondence --- the existing methods of the analytic bootstrap are applicable, but most of them fail to take advantage of the conformal manifold. Instead, they treat a fixed CFT irrespective of its deformations. New methods are needed to take advantage of the vast additional data encoded elsewhere in the conformal manifold, and to address questions about the manifold itself, such as whether there are universal properties of CFTs at infinite distance \cite{Perlmutter:2020buo}. See \cite{Behan:2017mwi,Chester:2021aun,Collier:2022emf} for recent works exploring conformal manifolds from the bootstrap perspective.

Massive deformations trigger renormalization group flows. Can the methods of Lorentzian CFT be extended to encompass these deformations? Is there a Lorentzian inversion formula for the data of effective field theories? What is the relation, if any, between the sign constraints obtained through the lightcone bootstrap, such as the ANEC, and the universal inequalities known as $C$-theorems that govern renormalization group flows?

The existence of moduli spaces of exact supersymmetric vacua gives rise to powerful tools for understanding the rich physics of strongly-coupled QFTs \cite{Seiberg:1994rs,Seiberg:1994aj}. What can the conformal bootstrap teach us about the moduli spaces of vacua? For example, how can we tell, purely from the CFT data, that a CFT possesses a nontrivial moduli space? See \cite{Beem:2014zpa} for a discussion in the context of $\mathcal{N}=2$ SCFTs.

\subsection*{Averaging and randomness}

Several independent directions, including condensed matter applications, the eigenstate thermalization hypothesis (ETH) \cite{PhysRevA.43.2046,Srednicki:1994mfb}, exciting recent developments in holographic duality, and hints from mathematics all suggest an important role for randomness in the conformal bootstrap. 

In 2019, it was discovered that two-dimensional gravity is holographically dual to random matrix theory \cite{Saad:2019lba}. This is fundamentally different from other known examples of holography because the boundary theory is not a single CFT, but an ensemble of quantum theories. Randomness, at some level, also appears to be an important feature of higher-dimensional black holes \cite{Cotler:2016fpe}. How does this manifest in the conformal bootstrap? Are there universal constraints on the statistics of CFT data that can be derived analytically? The answer is almost certainly yes, and this may provide a handle on complex, chaotic phenomena in CFT that are inaccessible to standard numerical and analytic methods.  Tauberian theorems that underlie the Cardy formula \cite{Cardy:1986ie} and the asymptotics of OPE coefficients \cite{Pappadopulo:2012jk} always involve some averaging \cite{Mukhametzhanov:2018zja,Mukhametzhanov:2019pzy}, so these two subjects are closely linked. Other bootstrap tools that are likely to play a significant role are the fusion kernel and, in 2d CFT, large-$c$ methods that reveal aspects of 3d gravity from a CFT perspective \cite{Hartman:2013mia,Fitzpatrick:2014vua,Asplund:2014coa,Collier:2018exn,Collier:2019weq}.  For initial steps in these directions, see e.g.\ \cite{Belin:2020hea,Belin:2020jxr,Belin:2021ibv,Anous:2021caj,Pollack:2020gfa,Cotler:2020ugk,Cotler:2020hgz,Heckman:2021vzx}.

Similar questions have been explored in the context of eigenstate thermalization, which is believed to govern the statistics of certain OPE coefficients responsible for linear response at finite temperature or density. If there are universal statistics in CFT accessible to the analytic bootstrap, what are their consequences for non-holographic theories like the 3D Ising model? Can the analytic bootstrap be applied to disordered systems relevant to condensed matter?

From the mathematics side, the hints of randomness come from spectral analysis and analytic number theory. See e.g.\ \cite{mazur2008finding} for a sampling of the relevant mathematics, and \cite{Benjamin:2021ygh,Collier:2022emf} for  recent discussions in the context of the bootstrap. Can a wider array of mathematical tools be applied to conformal field theory, and vice-versa? 

It was recently shown that aspects of holographic duality can be explained by averaging over the conformal manifold of CFTs in a two-dimensional toy model \cite{Maloney:2020nni,Afkhami-Jeddi:2020ezh} and then in ${\cal N}=4$ super-Yang Mills \cite{Collier:2022emf}. It has also been suggested that the randomness of black hole microstates is related to averaging over $N$ \cite{Schlenker:2022dyo}. Another possibility is to input some CFT data, such as the scaling dimensions and OPE coefficients of light operators, then choose a subset of bootstrap constraints and average over a `swampland' of CFT data that satisfy the constraints but do not necessarily have a UV completion. These ideas are developing rapidly in various directions; what are the best approaches to averaging in realistic theories of quantum gravity and higher-dimensional CFTs?

\subsection*{Closing the gap between analytics and numerics}

The analytic bootstrap, in its simplest form, determines asymptotic data at large spin and/or scaling dimension. It can be improved iteratively to extend its reach toward lighter operators. Numerical methods, meanwhile, effectively work upwards, starting with low-lying operators and producing more data for higher dimension operators as more constraints are included. This offers the tantalizing possibility that if these two approaches can be combined effectively, harnessing analytic results as input to numerics, it could lead to highly efficient methods to solve for CFT data throughout the spectrum. 

For example, in CFTs like the 3d Ising model, among the hundreds of scaling dimensions and OPE coefficients that can be found by semidefinite programming, most can be fit to analytics with the input of a few pieces of numerical data \cite{Simmons-Duffin:2016wlq,Liu:2020tpf,Caron-Huot:2020ouj,Atanasov:2022bpi}. This suggests that much of the computational power utilised by current methods may be spent on operators that are amenable to analytics. One of the issues with combining analytics and numerics is that error bars in analytic bootstrap results are currently not under precise control. At the same time, introducing errors into the setup of a numerical bootstrap computation can destroy the rigor of the approach. Improvements in the analysis of errors in large spin-expansions, and creative methods for taking those errors into account in numerics are needed, see \cite{Su:2022xnj} for recent work in this direction. 

In large-$N$ theories, standard numerical methods are less effective, but analytic techniques that take advantage of the $1/N$ expansion are more powerful. This is therefore another area where closer integration of analytics and numerics would be helpful.

It seems unlikely that there will ever be a fully analytic solution of chaotic CFTs. Nonetheless, it may be possible to find fast numerical algorithms powered by analytic bootstrap results that converge quickly for many observables. A near-term goal in this direction is to better understand the role of multi-twists in the analytic bootstrap, and to incorporate these into iterative methods to determine spectra through a combination of analytics and numerics.

\subsection*{Lorentzian signature and analytic structure of correlators}

Most of the analytic bootstrap methods discussed above require studying CFTs in Lorentzian signature. In principle,  Lorentzian constraints should follow from CFT axioms, which are usually formulated in Euclidean signature. However, Lorentzian signature packages important constraints in an accessible and useful way --- in particular by allowing one to formulate and explore the condition of causality \cite{Adams:2006sv,Camanho:2014apa,Hartman:2015lfa}. Relatedly, Lorentzian signature admits a much richer array of singularities than Euclidean signature, for example the lightcone, Regge, and ``bulk point" limits \cite{Gary:2009ae,Maldacena:2015iua,Iliesiu:2018fao,Dodelson:2020lal}. See \cite{Bercini:2021jti} for a recent bootstrap analysis of gauge theory correlators near a Lorentzian singularity, and 
\cite{Engelhardt:2016wgb,Engelhardt:2016crc,Hernandez-Cuenca:2020ppu} for a discussion of Lorentzian singularities in the context of bulk reconstruction. It was conjectured in \cite{Maldacena:2015iua} that Lorentzian singularities should be in one-to-one correspondence with position-space Landau diagrams. (Furthermore, emergent singularities in holographic CFTs should correspond to Landau diagrams in the bulk.) However, we are far from being able to verify that conjecture in $d>2$ dimensions using bootstrap methods. 

Many of the Lorentzian results discussed above (including the bootstrap proof of the ANEC \cite{Hartman:2016lgu}, the proof of commutativity of ANEC operators \cite{Kologlu:2019bco}, and arguments for the convergence of dispersive sum rules \cite{Caron-Huot:2020adz}) require technical assumptions about the behavior of CFT correlators under analytic continuation to the so-called ``second sheet," see \cite{Kravchuk:2021kwe} for a discussion. To place these results on a completely rigorous footing, it will be crucial to prove bounds on analytically-continued correlators, ideally with controlled expansions around Lorentzian singularities.  Recent progress in this direction was made in \cite{Kravchuk:2020scc,Qiao:2020bcs,Kravchuk:2021kwe}, which proved (Lorentzian) Wightman axioms for CFT four-point functions starting from Euclidean CFT axioms (bypassing technical assumptions in a famous theorem of Osterwalder and Schrader \cite{Osterwalder:1973dx,Osterwalder:1974tc}). Can old results in multi-dimensional complex analysis be applied fruitfully to this problem, or will new methods be needed?

Finally, it is an interesting problem to understand the analytic structure of CFT correlators in momentum space. See~\cite{Gillioz:2018mto,Gillioz:2019lgs,Gillioz:2020mdd,Meltzer:2021bmb,Gillioz:2021sce} for a sampling of momentum-space methods and results in CFT.

\subsection*{Quantum information, algebraic methods, and higher symmetries}

Many results of the analytic bootstrap are fundamentally consequences of locality. For example, the existence of double-twist operators reflects the fact that two operators inserted far apart will have little influence on each other, and so there must be composite primaries with this property. On the other hand, locality is not manifest in the bootstrap, and the operator language of CFT is quite different from other approaches to the physics of interacting quantum field theory where locality is built into the axioms. How does the bootstrap connect to these other approaches, such as algebraic QFT, tensor network-inspired methods, or topological quantum field theory? 

Approaches to QFT that make locality manifest also more naturally incorporate extended objects, higher symmetry groups, and anomalies.  It is an open question how to better use these structures in bootstrap calculations. In two dimensions, anomalies and line defects lead to refined modular bootstrap results, see e.g.\ \cite{Lin:2019kpn,Lin:2021udi,Grigoletto:2021zyv,Grigoletto:2021oho}. However, in higher-dimensions, bootstrap calculations of local operator spectra have so far only accessed anomalies that appear in local correlation functions. How do we improve bootstrap methods to explore the full structure of extended QFT?

Both algebraic QFT and tensor methods are intimately connected to quantum information theory. They rely on cutting manifolds into spatial subregions, i.e., across codimension-2 separating surfaces, and therefore the spatial organization of entanglement plays a central role. Connections to quantum gravity give another hint that quantum information has a larger part to play in the conformal bootstrap. One concrete goal in this area is to understand the relationship between the bootstrap derivation of the averaged null energy condition (ANEC) \cite{Hartman:2016lgu} and the information-theoretic derivation \cite{Faulkner:2016mzt}. Another target is to unify disparate approaches to $C$-theorems in various dimensions: How does the bootstrap derivation of the four-dimensional $a$-theorem \cite{Komargodski:2011vj} relate to the methods based on quantum entropy \cite{Casini:2017vbe} or holographic duality \cite{Myers:2010xs}?

\section{Outlook}

Developments over the past decade have demonstrated the surprising power of the basic principles of unitarity, causality and symmetry in constraining conformal field theories. At the same time, it seems likely that we have only been able to harness a small fraction of the full potential offered by these principles. Here, we have reviewed the main existing results and suggested several promising directions for future research. Whether or not our universe is the unique solution of the ultimate bootstrap constraints, it is clear that the conformal bootstrap will continue to illuminate the path towards a deeper understanding of quantum field theory and quantum gravity.

\section*{Acknowledgements}
We would like to thank Simon Caron-Huot for helpful discussions. TH is supported by the Simons Foundation and NSF grant PHY-2014071. DM acknowledges funding provided by Edward and Kiyomi Baird as well as the grant DE-SC0009988 from the U.S. Department of Energy.  DSD is supported by Simons Foundation grant 488657
(Simons Collaboration on the Nonperturbative Bootstrap) and a DOE Early Career Award
under grant no.\ DE-SC0019085. AZ received funding from the European Research Council (ERC) under the European Union’s Horizon 2020 research and innovation programme (grant agreement number 949077).

\bibliographystyle{JHEP}
\bibliography{refs}

\end{document}